\documentclass[twocolumn,showpacs,preprintnumbers,amsmath,amssymb,showkeys]{revtex4}
\usepackage{graphicx}
\usepackage{dcolumn}
\usepackage{bm}
\begin{document}
\title{Decay of the Mixed States}
\author{Yu.\,L.\,Bolotin}
\email{bolotin@kipt.kharkov.ua}
\author{ V.\,A.\,Cherkaskiy}
\author{G.\,I.\,Ivashkevych}
\affiliation{A.I.Akhiezer Institute for Theoretical Physics,
National Science Center "Kharkov Institute of Physics and
Technology", Akademicheskaya Str. 1, 61108 Kharkov, Ukraine}
\date{\today}
\begin{abstract}
We study the classical escape from local minima for 2d multi-well
Hamiltonian systems, realizing the mixed state. We show that escape
from such local minima has a diversity of principally new features,
representing an interesting topic for conceptual understanding of
chaotic dynamics and applications.
\end{abstract} \maketitle
The escape of trajectories (particles) from localized regions of
phase or configuration space has been an important topic in
dynamics, because it describes the decay phenomena of metastable
states in many branches of physics: chemical and nuclear reactions,
atomic ionization, nuclear fusion and so on. This problem has the
rich history. Almost a century ago, Sabine \cite{sabine} considered
the decay of sound in concert halls. Legrand and Sornette
\cite{legrand} have shown that this problem is equivalent to the
escape one: a small opening of width $\Delta$ for escape must be
identified with $\int\alpha(S)ds$, where $\alpha(s)$ is the
absorption coefficient at position $S$ of the container (billiard)
boundary, $\alpha(s)=1$ over the width of window and $\alpha(s)=0$
elsewhere. Szepfalusy and Tel \cite{tel} connected escape problem
with problem of chaotic scattering.

Exponential decay is a common property expected in strongly chaotic
classical systems \cite{bauer,alt,nemes}. Let us consider as an
example \cite{bauer} point particles bouncing elastically off the
walls in a rectangular box. The system is allowed to decay by
providing a small window in one of the box walls through which
particles can escape. As is well known, motion of particles in a
rectangular billiard is regular: two independent integrals of motion
are absolute values of momentum projection on the billiard walls.
The trajectories of particles become chaotic if a circular
scattering center is placed somewhere inside the box.

For the chaotic case simple consideration leads to the exponential
decay. The number of particles leaving per time interval is given by
\begin{equation}\label{n_t}\frac{dN}{dt}=\Delta\rho(t)\int d^2p\
\mathbf{pe}_n=-2\Delta\rho(t)p^2\delta p\end{equation} Here $p$ is
absolute value of the momentum, $\mathbf{e}_n$ is a unit vector
normal to the opening in the surface, and integration in momentum
space is taken over a circular ring with radius $p$ and
infinitesimal width $\delta p$. Function $\rho(t)$ is the phase
space density, which for ergodic motion is only a function of time.
In our case
\begin{equation}\label{rho_t}\rho(t)=\frac{N(t)}{2\pi p\delta
pA_c}\end{equation} where $A_c$ is total coordinate space area
available. Inserting (\ref{rho_t}) into (\ref{n_t}) yields
\begin{equation}\label{ntb}N(t)=N(0)e^{-\alpha t};\ \alpha=\frac{p\Delta}{\pi A_c}\end{equation}
Analytically calculated decay constant $\alpha$ is in a good
agreement with the graphically extracted value.

Exponential law at extremely long times turns into the power law
typical for decay of regular systems. One possible mechanism for
generation of power tails is the effect of "sticking" of the chaotic
orbits to outer boundaries of stability islands \cite{karney}, or a
very similar effect, connected with the existence of marginally
stable periodic or "bouncing ball" orbits. Although some qualitative
models, which show how the algebraic tail emerges, were introduced
in \cite{bunimovich}, no critical conditions for the distinct decay
laws were formulated in terms of the billiard geometrical
constrains. Experimental escape of cold atoms from a laser trap of
billiard type with a hole was studied in \cite{milner,friedman}.

Transition from the billiards to potential systems substantially
broadens number of possible applications of the escape problem, but
from another hand significantly complicates the problem. Of course,
the one-well case is the simplest one. Zhao and Du \cite{zhao_du}
reported a study on the escape rates near threshold of Henon-Heiles
potential
\begin{equation}\label{hh}
U_{HH}(x,y)=\frac{x^2+y^2}{2} + xy^2-\frac{x^3}{3}.
\end{equation}
Simulations performed by the authors show that the escape of
Henon-Heiles system at energy, slightly exceeding the saddle one,
follows exponential law similar to the chaotic billiard systems.
They derived an analytic formula for the escape rate as function of
energy. The derivation is based on the fact that the phase space of
the considered potential (as well as for the billiards) is
practically homogeneous near the saddle points. It should be noted
that in such case all trajectories with energy higher than the
saddle one leave the potential well in finite time. The only problem
to solve is to determine the probability of particle to escape from
the well in unit time interval.

In contrast to billiards, generic potential systems have essentially
inhomogeneous phase space structure. We intend to study the
particles escape from the local minima in the case when the phase
space contains macroscopically significant components of regular as
well as of chaotic type. Such possibility is realized in multi-well
potentials.

The principal peculiarity of the regularity-chaos transition in
multi-well potentials lies in the existence of different critical
energies for different local minima. It means that in such
potentials at one and the same energy in different local minima may
exist different dynamical regimes (either regular or chaotic). Such
kind of dynamics in multi-well potentials, when at some energy the
ratio of chaotic trajectories in certain local minimum significantly
differs from that ratio in other minima, is called the mixed state
\cite{paradigm}.

\begin{figure}
\includegraphics[width=0.23\textwidth,draft=false]{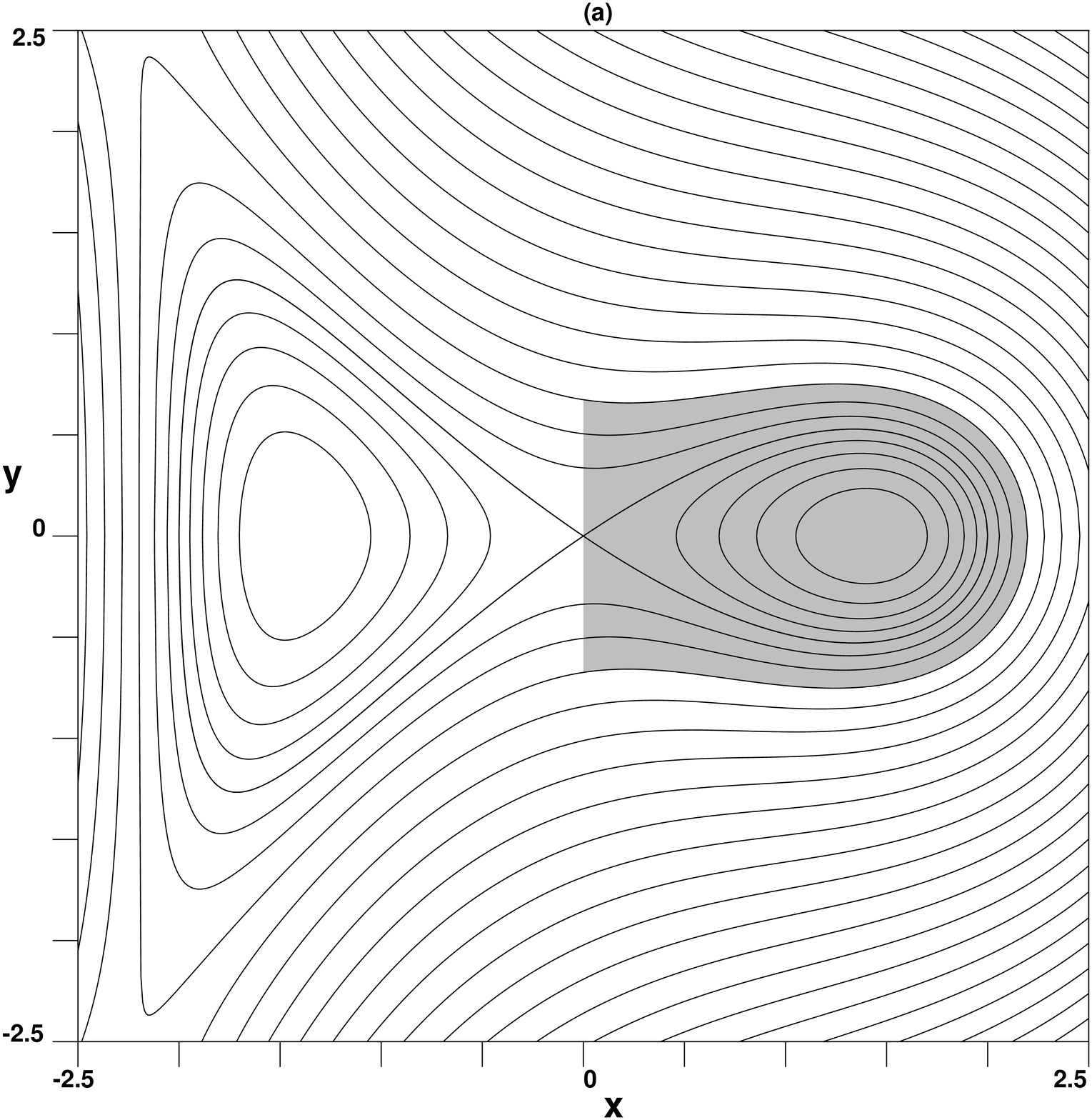}
\includegraphics[width=0.23\textwidth,draft=false]{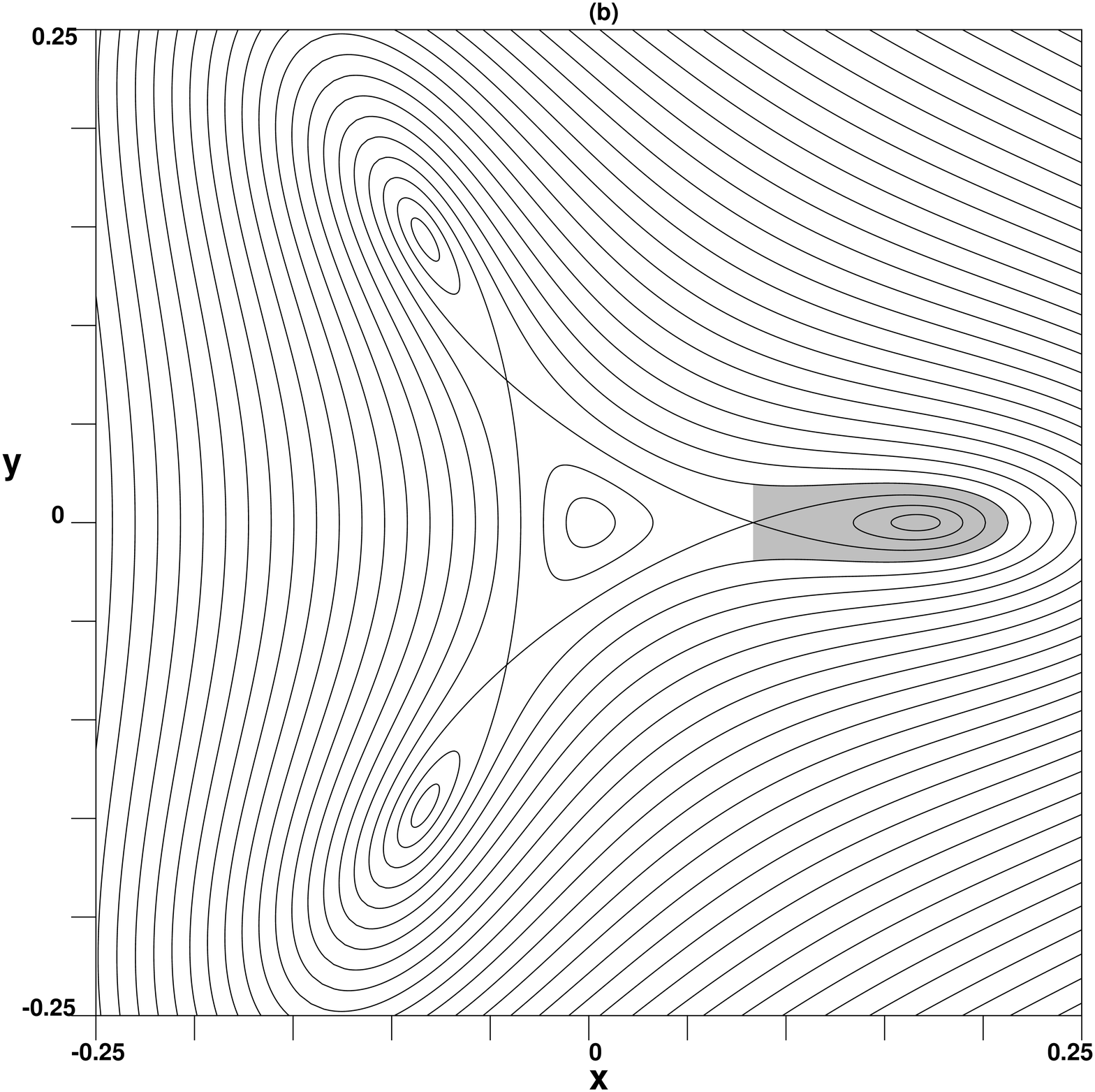}
\caption{\label{pes} Level lines for potential $D_5$ (\ref{d5}) (a)
and QO (\ref{qo}) (b). Gray color shows initial distribution of
particles for $E=2E_S$.}
\end{figure}

We demonstrate the mixed states on two representative examples: the
lower umbillic catastrophe $D_5$ potential \cite{gilmor}
\begin{equation}\label{d5}
U_{D_5}(x,y)=y^2(x+2a) + \left(\frac{x^2}{2}-1\right)^2
\end{equation}
for $a=1.1$ (fig.\ref{pes}.a) and the potential of quadrupole
oscillations (QO) of atomic nuclei \cite{mg}
\begin{equation}\label{qo}
U_{QO}(x,y)=\frac{x^2 + y^2}{2W} + xy^2-\frac{x^3}{3} +
\left(x^2+y^2\right)^2
\end{equation}
for $W=18$ (fig.\ref{pes}.b).
The potential $D_5$ (\ref{d5}) has only two local minima and three
saddles and it is the simplest potential, where the mixed state is
observed.

\begin{figure}
\includegraphics[width=0.11\textwidth,draft=false]{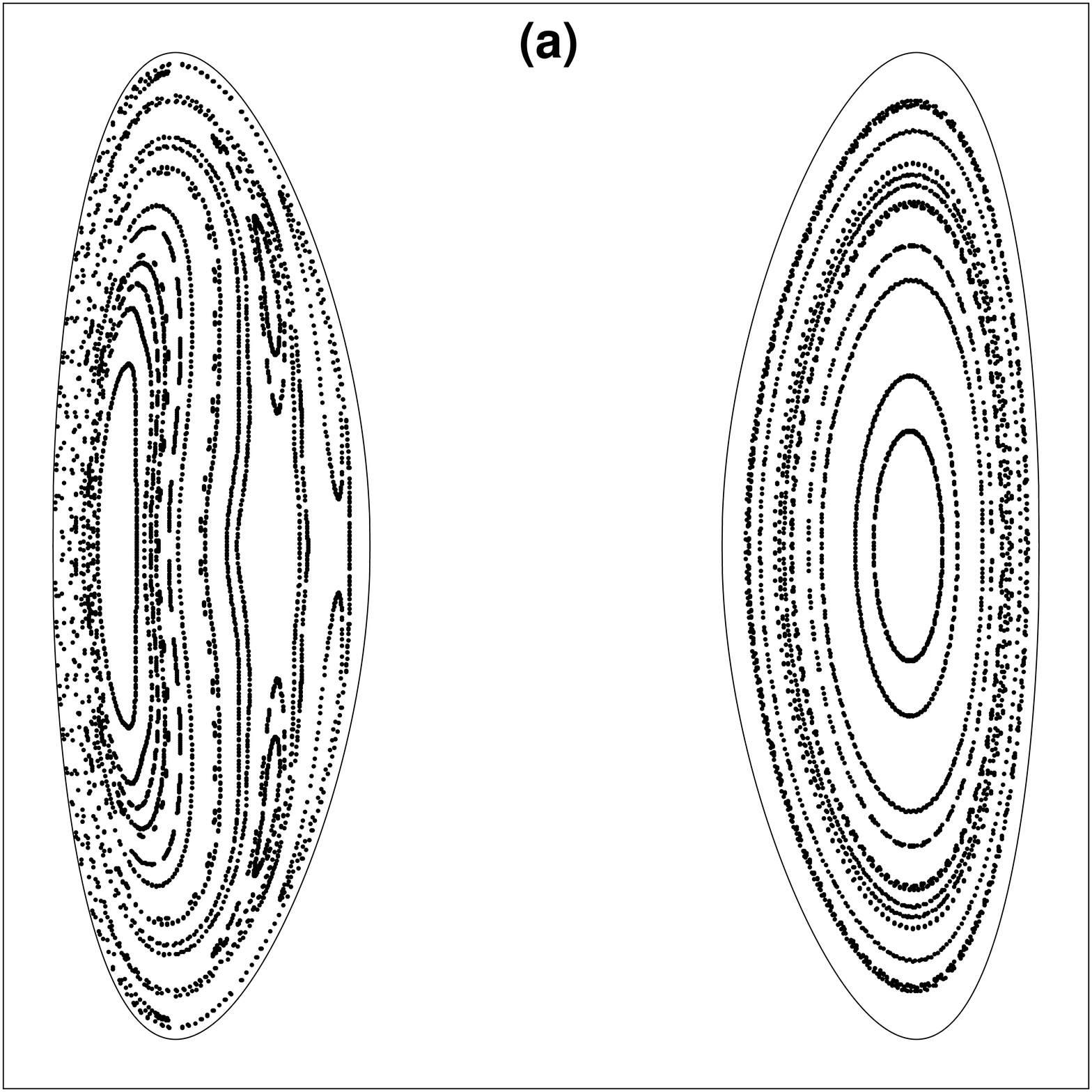}
\includegraphics[width=0.11\textwidth,draft=false]{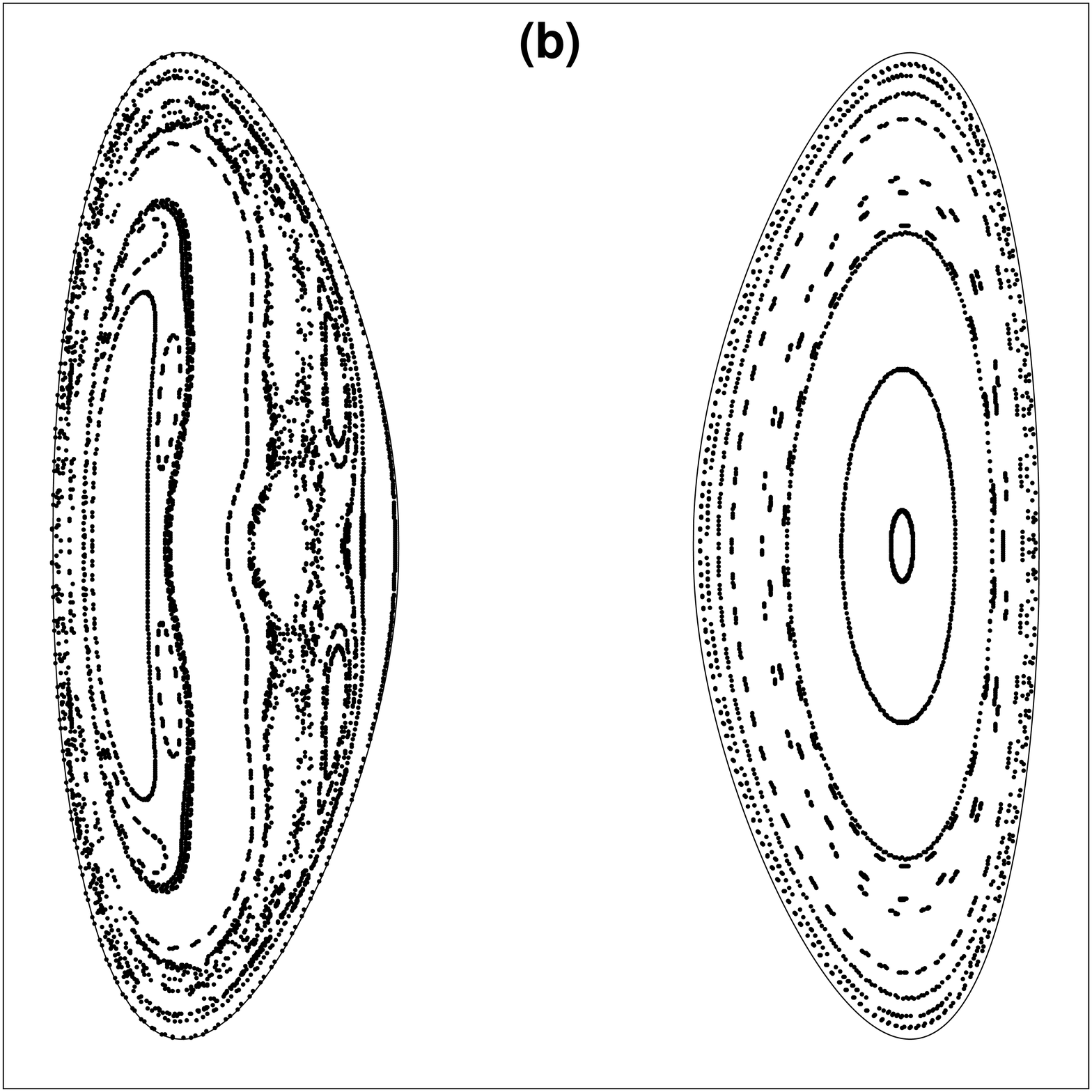}
\includegraphics[width=0.11\textwidth,draft=false]{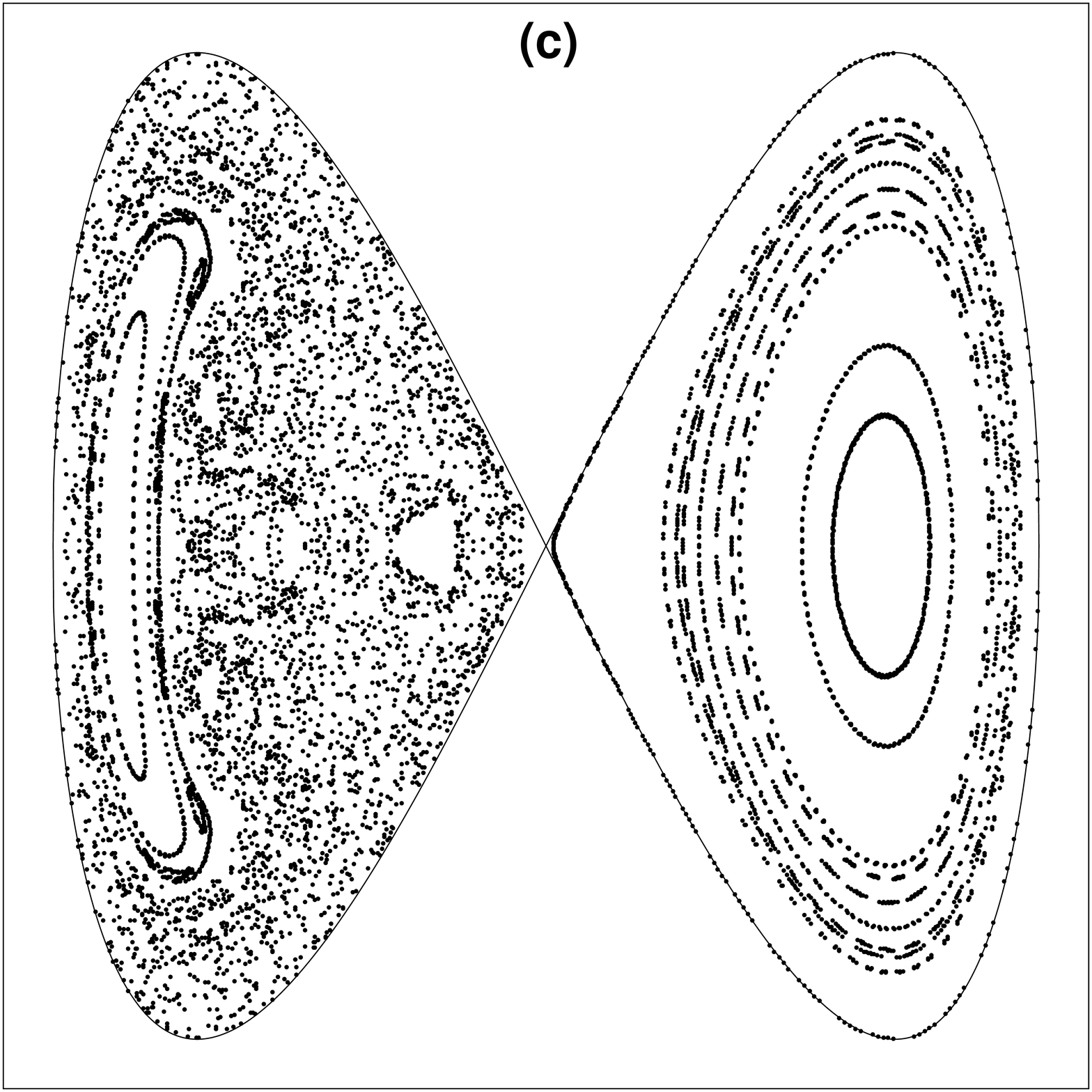}
\includegraphics[width=0.11\textwidth,draft=false]{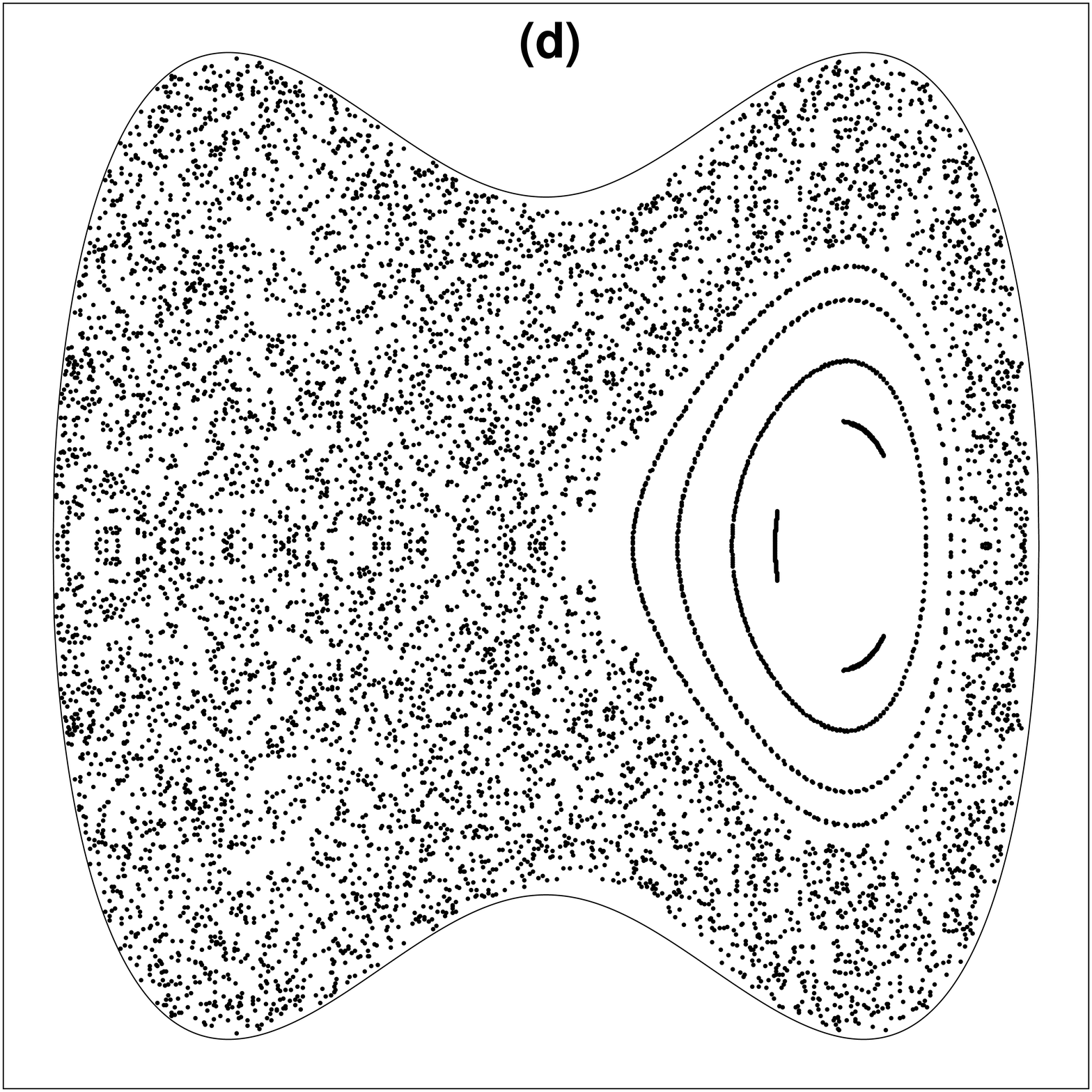}
\includegraphics[width=0.11\textwidth,draft=false]{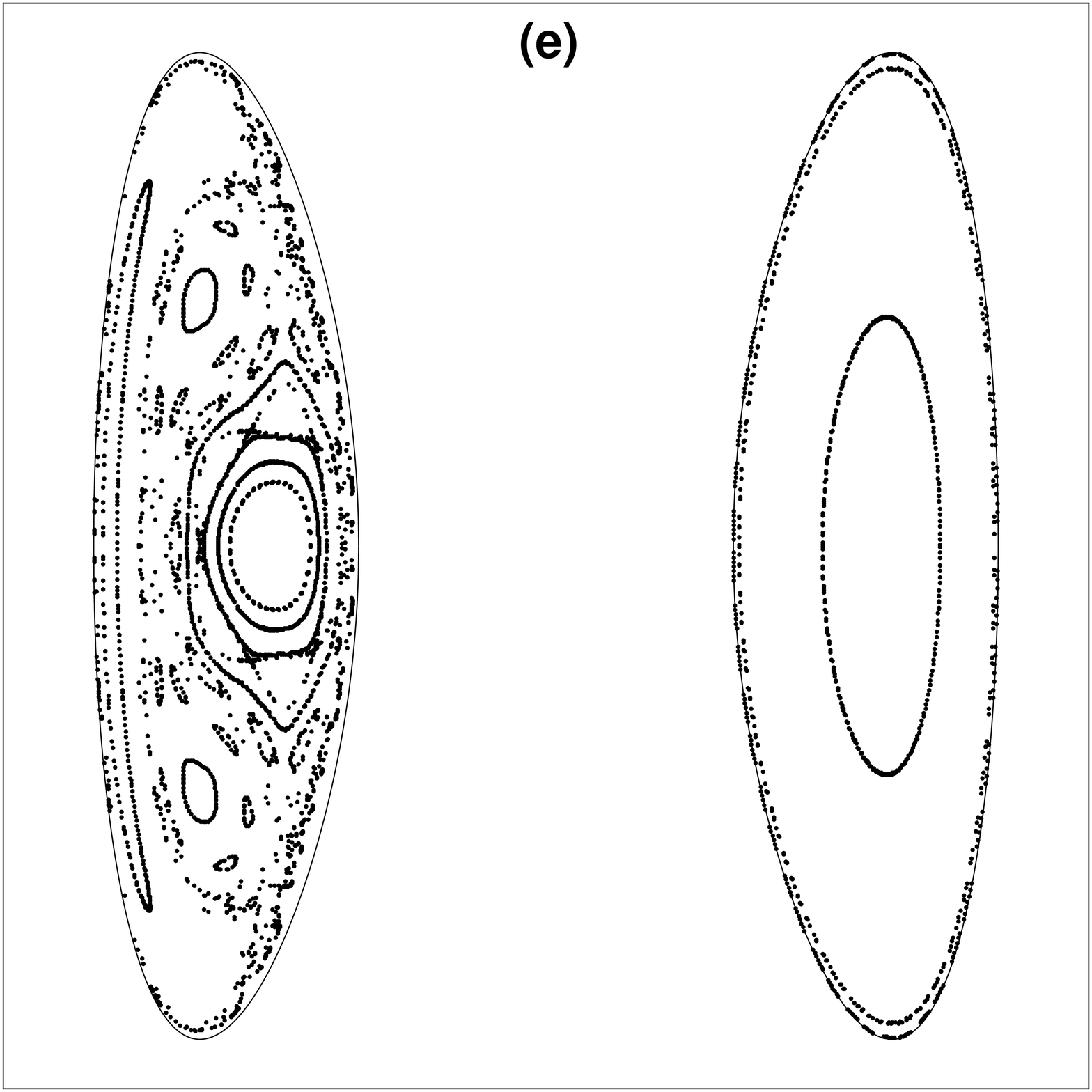}
\includegraphics[width=0.11\textwidth,draft=false]{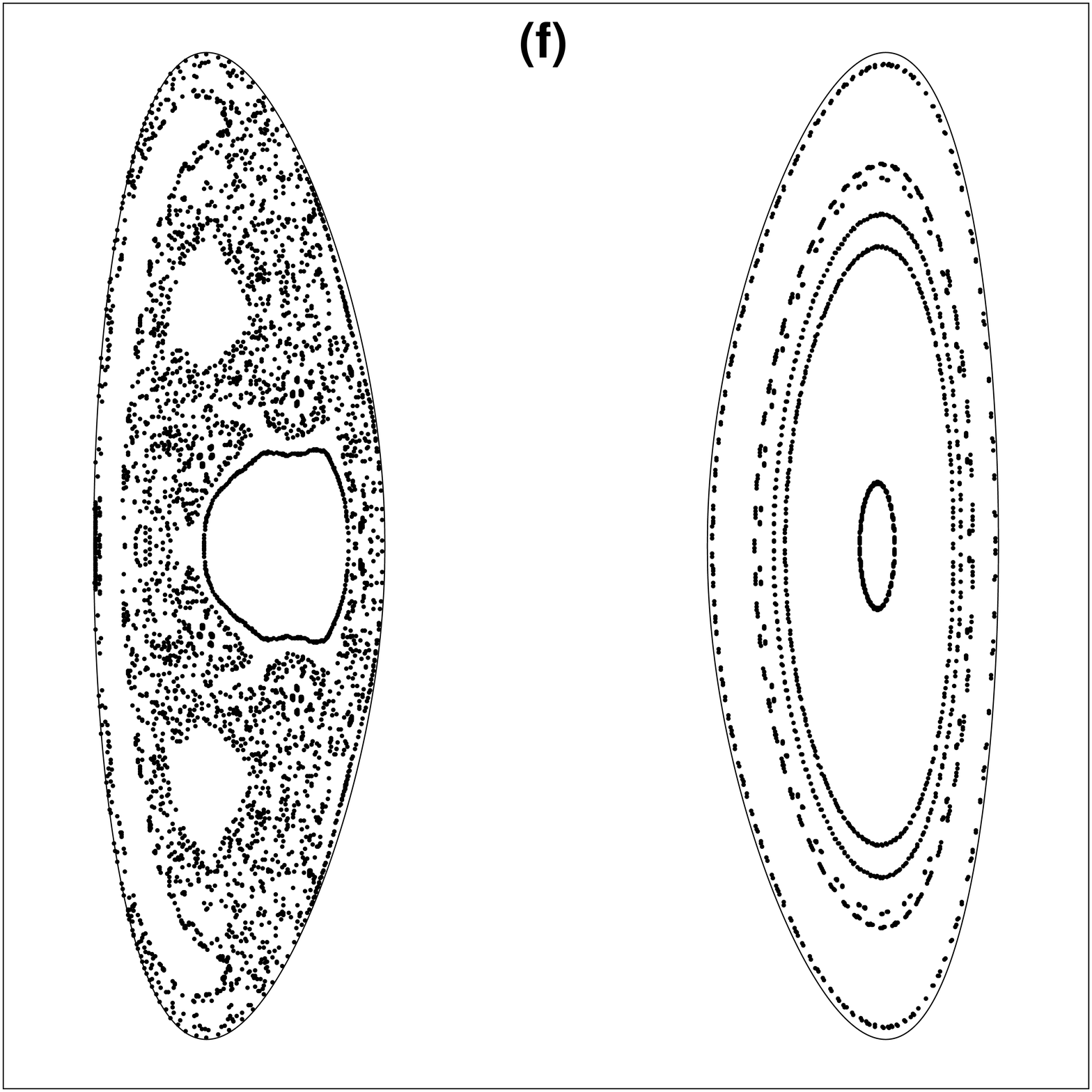}
\includegraphics[width=0.11\textwidth,draft=false]{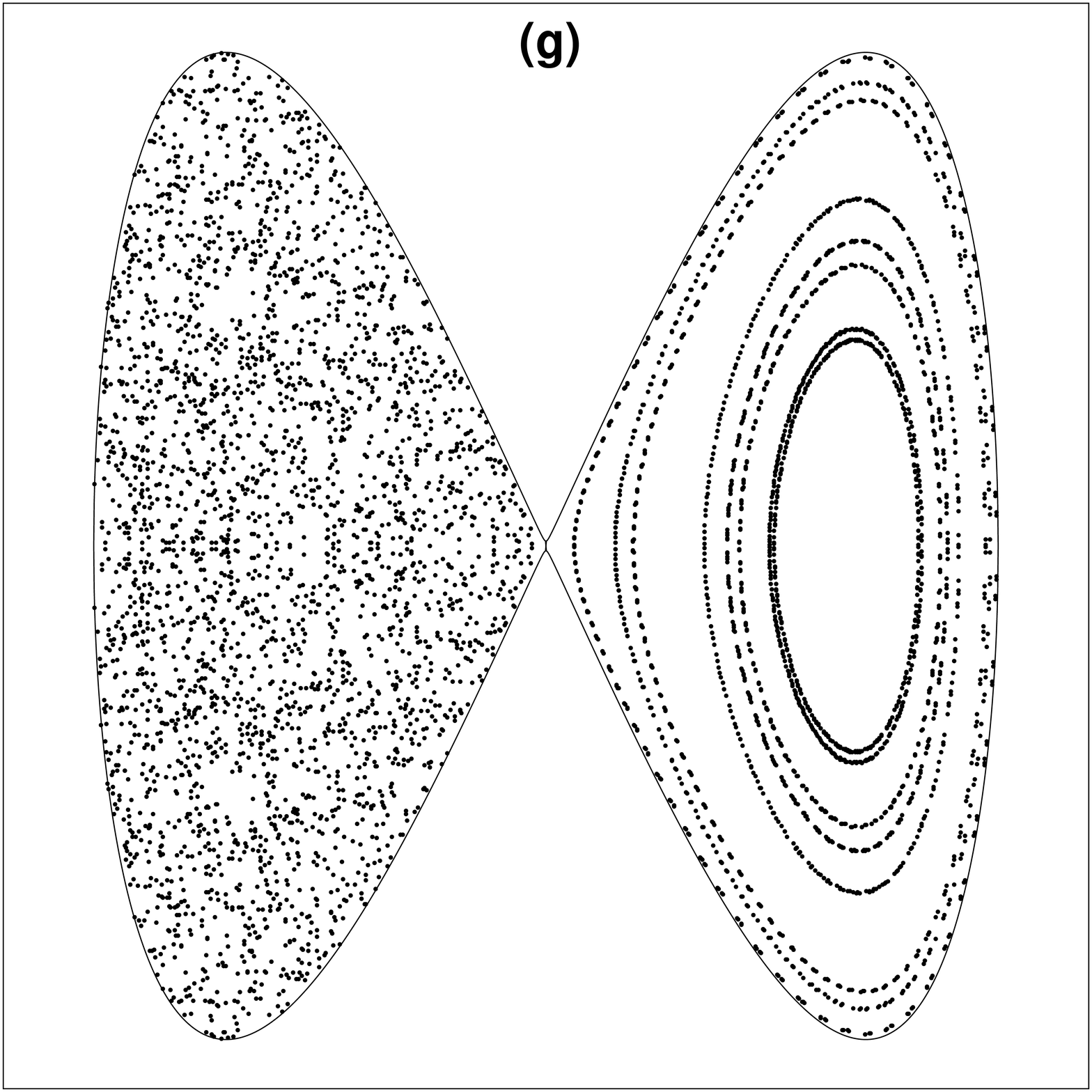}
\includegraphics[width=0.11\textwidth,draft=false]{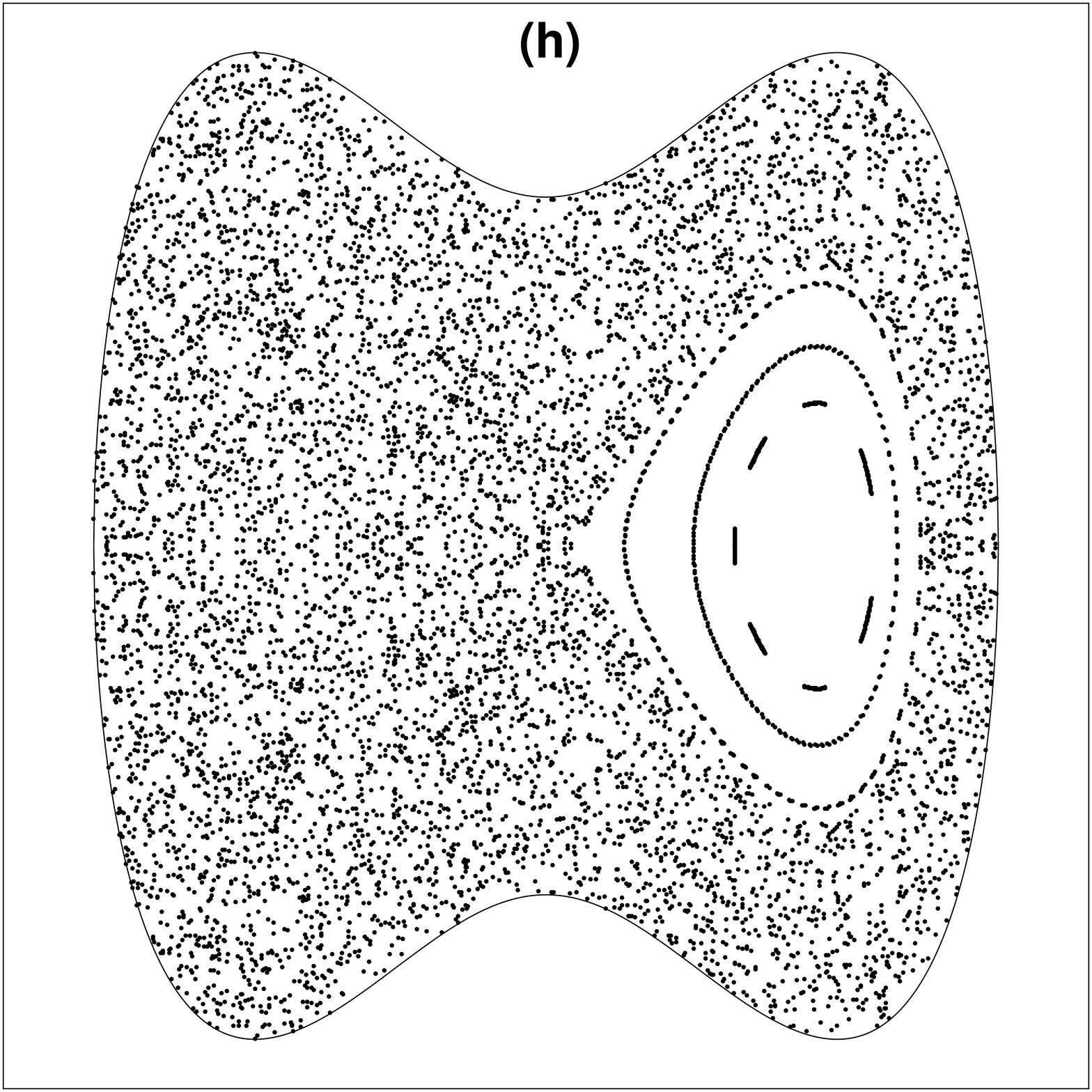}
\caption{\label{pss} Poncar\'e surfaces of section $y=0$ in the
$(x,p_x)$ plane for the potentials $D_5$ (\ref{d5}) (a-d) and QO
(\ref{qo}) (e-h) for different energies (from left to right):
$E\lesssim E_S$, $E\gtrsim E_S$, $E=E_S$, $E=2E_S$. Solid line
delimits the classically allowed region of phase space.}
\end{figure}

Fig.\ref{pss} presents the Poincar\'e sections for different
energies, demonstrating evolution of dynamics in different local
minima. At low energies motion has well-marked quasiperiodic
character for both minima (fig.\ref{pss}.a,e). As energy grows,
gradual regularity-to-chaos transition is observed. However changes
in features of the trajectories, localized in certain minima, are
sharply distinct. For the left minimum, already at about half saddle
energy, significant fraction of the trajectories becomes chaotic
(fig.\ref{pss}.b,f), and at saddle energy practically all initial
conditions produce chaotic trajectories (fig.\ref{pss}.c,g). In
right minimum under the same conditions motion remains quasiperiodic
up to the saddle energy (further we will call it "regular local
minimum" for simplicity). Moreover, at energies significantly higher
than the saddle energy (see fig.\ref{pss}), the phase space
structure preserves division on chaotic and regular components
(fig.\ref{pss}.d,h). The latter is localized in the part of the
configuration space which corresponds to regular motion at energies
below the saddle.
Earlier we have shown that the mixed state opens new possibilities
for investigations of quantum signatures of classical stochasticity
\cite{sm1,sm2}. Aim of the present work is to study the classical
escape from separated local minima, realizing the mixed state. We
show, that escape from such local minima has all the above mentioned
properties of decay of chaotic systems, and also a diversity of
principally new features, representing an interesting topic for
conceptual understanding of chaotic dynamics, and for the
applications as well. We are interested only in the "first passage"
effects, leaving aside the problem of dynamical equilibrium setup
for the finite motion (for example, in $QO$  potential). In is
important to stress that though we study the process of escape from
a concrete local minimum, the over-barrier in the case of mixed
state has a specific memory: general phase space structure at
super-saddle energies is determined by the characteristics of motion
in all other local minima.

We carried out numerical simulation and analytical estimates of
trajectories escape in the potentials $D_5$ and $QO$ through the
hole over the saddle point. Results of the escape problem for
systems with multi-component phase volume (regular and chaotic
components) essentially depend on choice of ensembles of initial
conditions for dynamical variables. Fig.\ref{n_t2} presents the
normalized particle number $N(t)/N(t=0)$ for $10^6$ initial
conditions, uniformly distributed inside the right minimum in the
potential $D_5\ (x>0)$ and peripheral minimum in the $QO$ potential
$(x>1/12)$ together with the typical trajectories and Poincare
sections. The results for different potentials are evidently similar
and have such characteristic features:
\begin{figure}
\includegraphics[width=0.5\textwidth,draft=false]{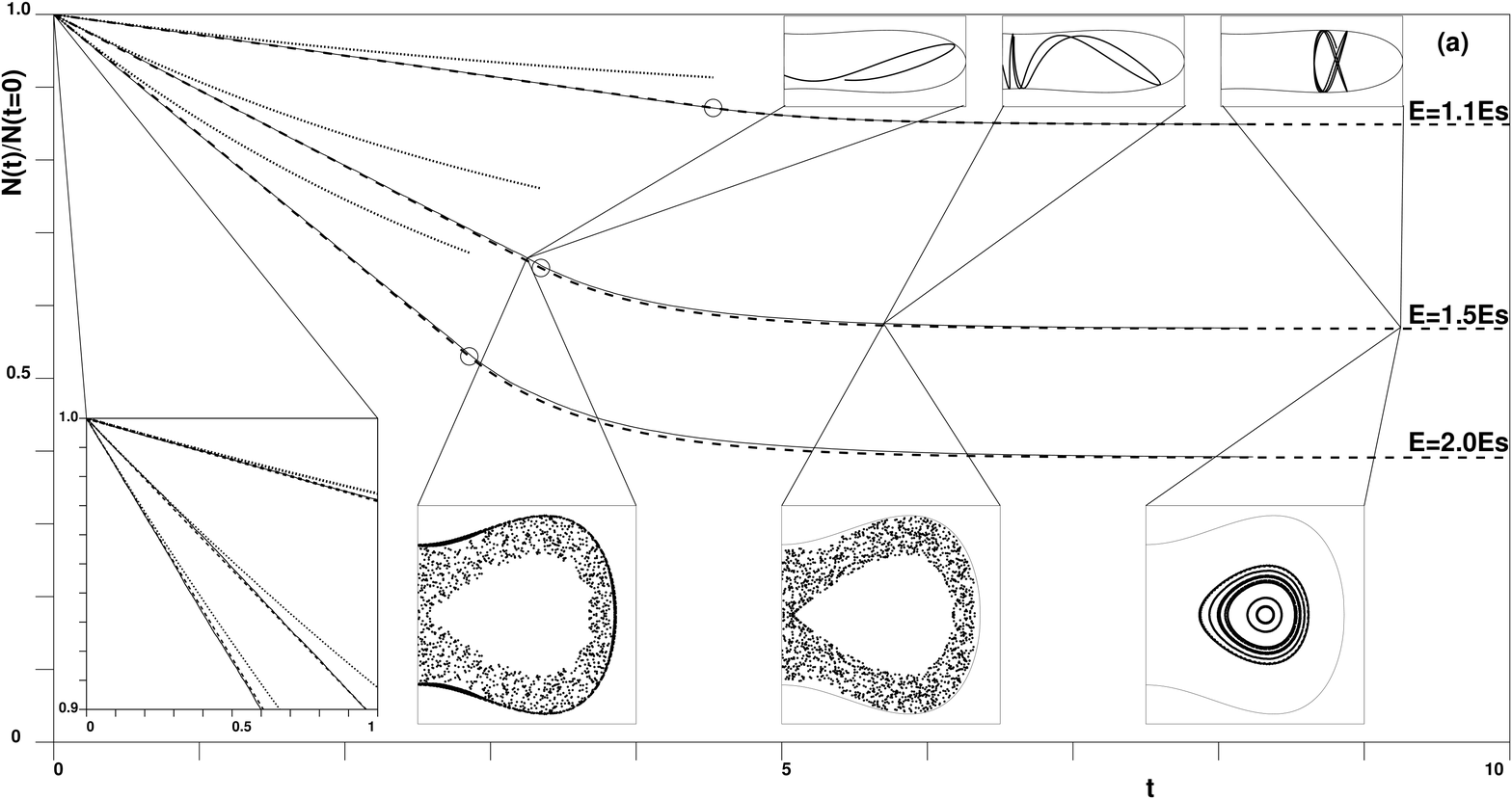}
\includegraphics[width=0.5\textwidth,draft=false]{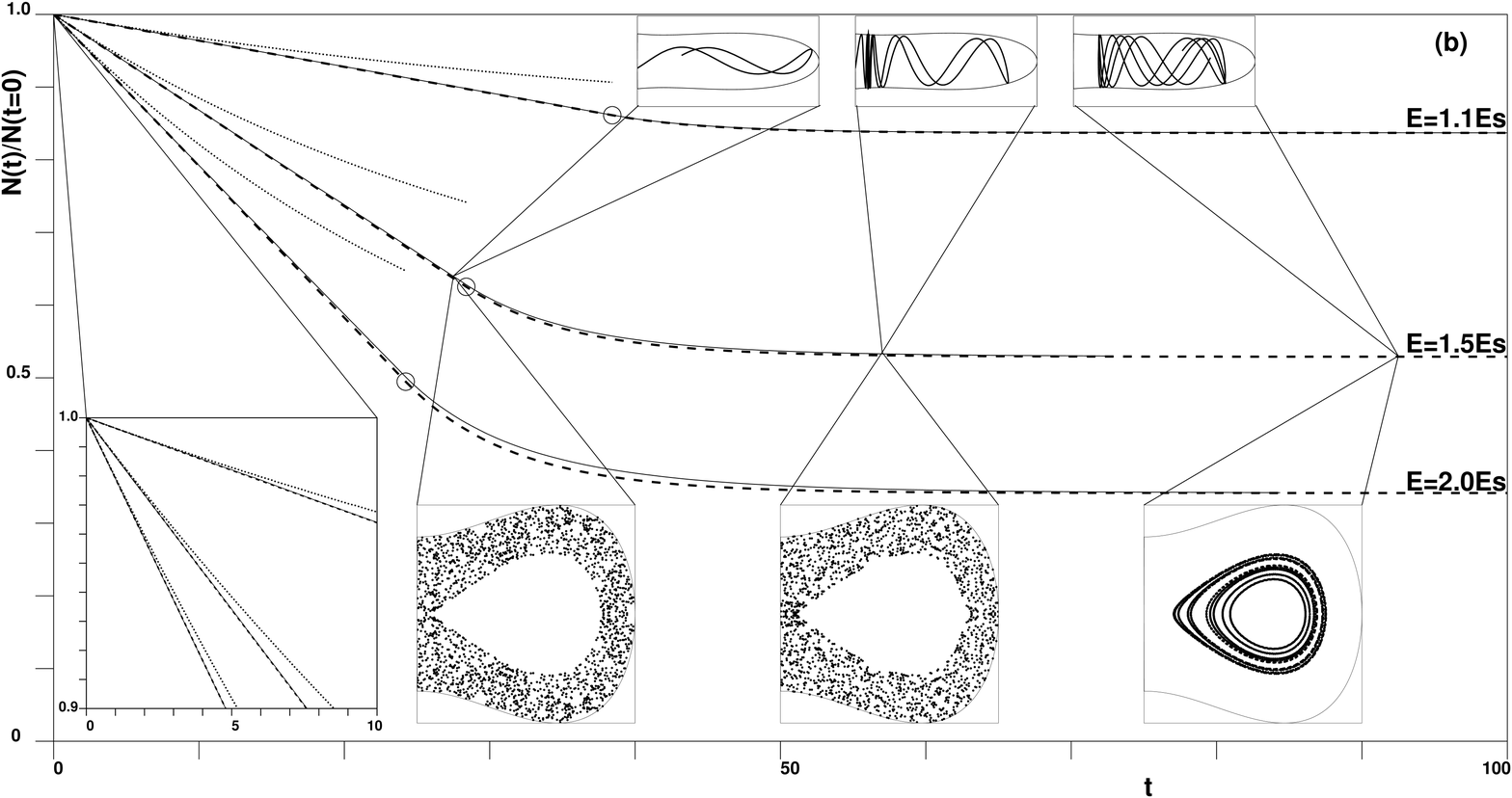}
\caption{\label{n_t2} Decay law for mixed states in the $D_5$ (a)
and $QO$ (b) potentials. Solid lines --- numerical simulation for
$E/E_S = 1.1, 1.5, 2.0$; dotted and dashed lines
--- theoretically obtained exponential and linear decay laws
respectively, zoomed on the inset figure in the lower left corners.
Other inset figures represent the typical trajectories and
Poincar\'e sections for the three different types of initial
conditions: linearly escaping, exponentially escaping and
non-escaping. Circles show the joining points between the linear and
the exponential decay laws at critical time $t=\tau$.}
\end{figure}
\begin{itemize}
\item  At times $t\rightarrow\infty$ decay law saturates at \[N(t\rightarrow\infty)=\rho^{(ne)}N_0\] where $\rho^{(ne)}$ is
equal to relative phase volume of "never-escaping" trajectories,
which represent the regular trajectories, completely localized
inside the considered minimum. All such trajectories, therefore,
have infinite escaping times.
\item For $t>\tau(E)$ the decay law
has exponential form \begin{equation}\label{exp} N(t)/N_0 =
\rho^{(ne)} + \rho^{(e)}e^{- \alpha^{(e)} (t-\tau)}\end{equation}
where $\rho^{(e)}$ represents relative number of exponentially
escaping particles.
\item For $t<\tau(E)$ the decay law is linear
\begin{equation}\label{lin}N(t)/N_0 = 1 - \alpha^{(l)}
t.\end{equation}
\end{itemize}

We should stress, that (\ref{lin}) is in no way a linear
approximation of (\ref{exp}) for small $t$: in general
$\rho^{(ne)}+\rho^{(e)}e^{-\alpha^{(e)}\tau}\ne1$ and
$\alpha^{(e)}\rho^{(e)}e^{-\alpha^{(e)}\tau}\ne\alpha^{(l)}$.
Instead, from the condition of smooth joining of curves (\ref{lin})
and (\ref{exp}) in the transition point $t=\tau$ we obtain
\[\begin{array}{c}
\alpha^{(e)} = \alpha^{(l)}/\rho^{(e)}\\
\rho^{(e)} = 1 - \rho^{(ne)} - \rho^{(l)}\end{array}\] where
$\rho^{(l)}=\alpha^{(l)}\tau$ is the relative number of linearly
escaping particles. Moreover, already on time scales
$t\lesssim\tau(E)$ linear decay law (\ref{lin}) is apparently
different from its exponential analogue \[N(t)/N(t=0)=\rho^{(ne)} +
(1-\rho^{(ne)})e^{(- \alpha^{(l)} t)}\] (see the inset on
fig.\ref{n_t2}).

As one can see from the inset Poincar\'e section on fig.\ref{n_t2}
both the chaotic and regular trajectories contribute to linear
escaping regime (\ref{lin}), because for sufficiently small times
$t<\tau$ chaotic and regular motions are not yet distinguishable. Up
to transient time $t=\tau$ all quasi-one-dimensional regular
trajectories, oriented along the $x$-axis, already escape and for
$t>\tau$ the escape of remaining chaotic particles follows
exponential law (\ref{exp}). The particles escaping the last show
already mentioned sticking phenomenon (see the inset Poincar\'e
sections on fig.\ref{n_t2}).

The transient time $\tau(E)$ in fact coincides with the passage time
of the longest one-dimensional path from the opening to the opposite
wall of the potential well and back (see fig.\ref{n_t2}). For the
potentials $D_5$ and $QO$ corresponding theoretical estimates read
(we assumed $m=1$) \begin{equation*}\begin{array}{c} \tau_{D_5}(E) =
2\int\limits_{0}^{\sqrt{2(1+\sqrt{E})}}\frac{dx}{|p|}=
\frac{\sqrt{2}}{E^{\frac 1 4}} K\left(\sqrt{\frac{1 + \frac{1}{\sqrt E}}{2}}\right)\\
\tau_{QO}(E) = 12\left(\frac{E_S}{E}\right)^{\frac 1 4}
K\left(\sqrt{\frac{1 + \sqrt\frac{E_S}{E}}{2}}\right) =
6\sqrt{2}\tau_{D_5}(\frac{E}{E_S})
\end{array}\end{equation*} where $K(k)$ is the complete elliptic integral of
the first kind and $E_S=1/12^4$ is the saddle energy in the $QO$
potential for $W=18$ (for the $D_5$ potential $E_S=1$).

Theoretical estimates for the escape rate were obtained by averaging
the escape probability over the opening \cite{zhao_du}:
\begin{equation*} \alpha(E) = \rho(E)\int\limits_{x=x_S} dy
\int\limits_{-\frac \pi 2}^{\frac \pi 2}d\theta |p|\cos\theta,
\end{equation*}
where $x_S$ is the coordinate of the saddle point and $\rho(E)$ is
the normalized particle density: \begin{equation*}
\rho(E)=\frac{1}{2\pi A(E)},
\end{equation*}
where $A(E)$ denotes area of the classically allowed region inside
the well: \[A(E)=\int\limits_{x>x_S} dx dy
\Theta\left(E-U(x,y)\right).\] Such $\rho(E)$ corresponds to uniform
distribution of initial conditions on the energy surface
$H(\mathbf{p,q})=E$.

For the potentials $D_5$ and $QO$ the explicit formulae are the
following:
\begin{eqnarray*}A_{D_5}(E)=2\int\limits_0^{\sqrt{2(1+\sqrt{E})}} dx
\sqrt{\frac{E-\left(\frac{x^2}{2}-1\right)^2}{x+2a}}\\
A_{QO}(E)=\frac{1}{72}\int\limits_0^{\sqrt{1+\sqrt{\frac{E}{E_S}}}}
d\xi \sqrt{(\xi+4)^2-7} \times \\
\times \sqrt{\sqrt{1+\frac{\frac{E}{E_S}
-\left(\xi^2-1\right)^2}{[(\xi+4)^2-7]^2}}-1}\end{eqnarray*} where
$\xi=\sqrt[4]{E_S}(x-x_S)$.

Finally, the general expression for the escape rate is
\begin{equation*} \alpha(E) = \frac{\langle p\rangle}{\pi A(E)}.
\end{equation*}
For our case $\langle p\rangle=\int\limits_{x=x_S} dy |p|$. In the
case of billiards with small opening $p=const\Rightarrow\langle
p\rangle=p\Delta$, and we recover expression (\ref{ntb}). For the
potentials $D_5$ and $QO$ we get the results in closed form:
\begin{equation*} \alpha_{D_5}(E) =
\frac{E-1}{2\sqrt{a}A_{D_5}(E)}\end{equation*}
\[\alpha_{QO}(E) = \frac{\sqrt[4]{\varepsilon}}{12\pi A_{QO}(E)}\times\]
\[\times\left\{\left(16\sqrt{\varepsilon}+1\right)K\left(\sqrt{\frac{1-\frac{1}{16\sqrt{\varepsilon}}}{2}}\right)
-2E\left(\sqrt{\frac{1-\frac{1}{16\sqrt{\varepsilon}}}{2}}\right)\right\}\]
where $\varepsilon=E-E_S+1/256$, $K(k)$ and $E(k)$ are the complete
elliptic integral of the first and second kind respectively.

In order to obtain $\alpha^{(l)}(E)$  we correct $A(E)$ subtracting
the relative phase space occupied by the non-escaping particles
\begin{equation*}A^{(l)}(E)=A(E)(1-\rho^{(ne)})\Rightarrow
\alpha^{(l)}(E)=\frac{\alpha(E)}{1-\rho^{(ne)}}\end{equation*}
Fig.\ref{n_t2} demonstrates good agreement between our theoretical
and numerical results for wide energy range.

\begin{figure}
\includegraphics[width=0.5\textwidth,draft=false]{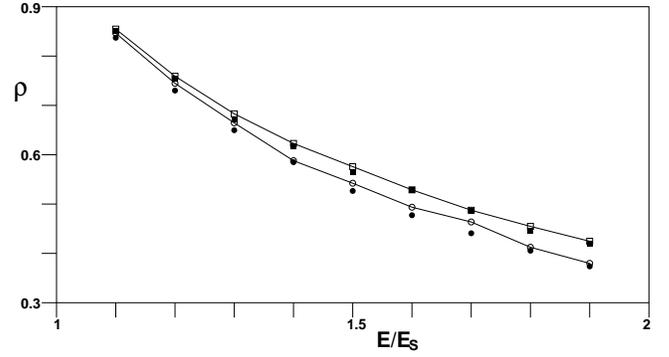}
\caption{\label{rho_e} Correlations between relative area of the
stability island $\rho^{(si)}(E)$ and non-escaping particles ratio
$\rho^{(ne)}$: empty squares -- $\rho^{(si)}_{D_5}(E)$, filled
squares -- $\rho^{(ne)}_{D_5}(E)$, empty circles --
$\rho^{(si)}_{QO}(E)$, filled circles -- $\rho^{(ne)}_{QO}(E)$.}
\end{figure}

Fraction of the non-escaping particles $\rho^{(ne)}$ coincides with
the relative phase space volume of trajectories, localized in the
regular minimum, which may be well estimated by the relative area of
the stability island $\rho^{(si)}$ on the Poincar\'e section
(fig.\ref{n_t2}). Calculation of relative area of regular island in
the Poincar\'e section was performed by the following scheme. First,
the island boundary was determined by numerical integration of
equations of motion and then the interior area was calculated.
Further the obtained area was divided on the entire area of
classically allowed motion, defined by the conditions $x>0$ and
$p^2>0$. While the phase volume itself is 4-dimensional, the
stability island in Poincar\'e section is 2-dimensional, so we
cannot expect absolute coincidence of the corresponding measures.
However, the calculations show very close correspondence between
them (see fig.\ref{rho_e}). Therefore, numerical analysis of
Poincare sections together with our theoretical results gives all
information necessary to predict the escape dynamics in independent
way.

In summary, we have considered classical escape from separated local
minima in two representative $2D$ multi-well potentials, realizing
the mixed state. We have found that escape from regular minima
contains a number of new features. The most important among them are
the following: \begin{enumerate} \item Decay law saturates at long
time ranges. \item On small time scales  there exists a linear
segment, which is not connected with linear approximation to the
exponential decay law, observed in chaotic systems with homogeneous
phase space. \item Fraction of particles, remaining in the well, is
determined by relative phase volume of the regular component, which
in its turn monotonically decreases with growth of energy.
\end{enumerate} It was shown that the linear segment of the decay law is
generated by the quasi-one-dimensional trajectories, oriented
perpendicular to the opening, and the transient time of the
linear-to-exponential regime lies in perfect agreement with the
analytical estimates.

We should note that we devote main attention to escape from the
regular local minima because the specifics of the mixed state
manifests only in them. However let us remind that in the case of
mixed state the phase space structure at super-saddle energies is
determined by dynamical characteristics in different local minima of
whole potential energy surface.

Above mentioned peculiarities of the escape problem may found
practical application for extraction of required particle number
from atomic traps. Changing energy of particles trapped inside the
regular minimum, we can extract from the trap any required number of
particles. Problem of particle energy changing in the potential well
may be solved by introduction of small dissipation. Obtained results
may present an interest also for description of induced nuclear
fission in the case of double-humped fission barrier. Revealed
peculiarities must manifest also in over-barrier dynamics of wave
packets, initially localized in the regular minima.

V.A.Cherkaskiy was supported by grant n.50-2007 of National Academy
of Science of Ukraine.
\bibliography{escape}
\end{document}